\documentclass[aps,prl,reprint,longbibliography,nobalancelastpage]{revtex4-2}

\usepackage{amsmath,amssymb}
\usepackage{graphicx}
\usepackage{xcolor}
\usepackage{mathrsfs}
\usepackage{stmaryrd}
\usepackage{hyperref}
\usepackage{enumitem}
\usepackage{accents}
\hypersetup{colorlinks,linkcolor={blue!80!black},citecolor={blue!80!black},urlcolor={blue!80!black}}

\renewcommand{\vec}[1]{\boldsymbol{#1}}

\renewcommand{\pi}{\uppi}

\DeclareMathAlphabet{\mathcal}{OMS}{cmsy}{m}{n}
\DeclareMathAlphabet{\mathcalbf}{OMS}{cmsy}{b}{n}
\DeclareMathAlphabet{\mathbfsfit}{\encodingdefault}{\sfdefault}{b}{it}
\DeclareMathAlphabet{\mathbfsf}{\encodingdefault}{\sfdefault}{b}{n}
\newcommand{\mat}[1]{\mathsf{#1}}

\newcommand{\figref}[2]{[Fig.~\hyperref[#1]{\ref*{#1}(#2)}]}
\newcommand{\figrefi}[2]{[Fig.~\hyperref[#1]{\ref*{#1}(#2)}, inset]}
\newcommand{\textfigref}[2]{Fig.~\hyperref[#1]{\ref*{#1}(#2)}}
\newcommand{\wholefigref}[1]{(Fig.~\ref{#1})}
\newcommand{\wholefigrefi}[1]{(Fig.~\ref{#1}, inset)}
\newcommand{\textwholefigref}[1]{Fig.~\ref{#1}}
\newcommand{\citeref}[1]{Ref.~\cite{#1}}

\renewcommand{\leq}{\leqslant}
\renewcommand{\geq}{\geqslant}

\begin{document}

\title{Subpopulations and Stability in Microbial Communities}
\author{Pierre A. Haas}
\email{P.A.Haas@damtp.cam.ac.uk}
\author{Nuno M. Oliveira}
\email{N.M.Oliveira@damtp.cam.ac.uk}
\author{Raymond E. Goldstein}
\email{R.E.Goldstein@damtp.cam.ac.uk}
\affiliation{Department of Applied Mathematics and Theoretical Physics, Centre for Mathematical Sciences, \\ University of Cambridge, 
Wilberforce Road, Cambridge CB3 0WA, United Kingdom}
\date{\today}%
\begin{abstract}
In microbial communities, each species often has multiple, distinct phenotypes, but studies of ecological stability have largely ignored this subpopulation structure. Here, we show that such implicit averaging over phenotypes leads to incorrect linear stability results. We then analyze the effect of phenotypic switching in detail in an asymptotic limit and partly overturn classical stability paradigms: abundant phenotypic variation is linearly destabilizing but, surprisingly, a rare phenotype such as bacterial persisters has a stabilizing effect. Finally, we extend these results by showing how phenotypic variation modifies the stability of the system to large perturbations such as antibiotic treatments.
\end{abstract}

\maketitle
Over forty years ago, May suggested that equilibria of large ecological communities are overwhelmingly likely to be linearly unstable~\cite{may72}.  
His approach did not specify the details of the dynamical system that describes the full population dynamics, but rather assumed that the linearized dynamics near the fixed point were represented by a \emph{random} Jacobian matrix.  
Invoking results from random matrix theory, he concluded that unstable eigenvalues are more likely to arise as the number of interacting species increases. Actual large ecological communities certainly seem stable, and a 
major research theme in theoretical ecology has been to identify
those features of the population dynamics that stabilize them~\cite{roberts74,allesina12,grilli16,grilli17,butler18,servan18}. 

Recent advances in our understanding of large natural microbial communities such as the human microbiome have emphasized the important link between stability 
and function: adult individuals typically carry the same microbiome composition
for long periods of time and disturbances thereof are often
associated with disease~\cite{lozupone12,faith13,coyte15}. 
Moreover, while genetically identical organisms may exhibit
different phenotypes \cite{kussell05b,smits06,avery06,dubnau06} and
despite the known ecological importance of phenotypic variation~\cite{turcotte16}, studies of stability have largely ignored the existence of such subpopulations within species.
Most models are therefore \emph{implicit averages} over 
subpopulations.

We show here that this averaging yields incorrect 
stability results. With stochastic switching between phenotypes as an example 
of subpopulation structure, we show that while multiple abundant phenotypes 
are destabilizing, a rare phenotype
can be stabilizing. This surprising result partly overturns May's paradigm, 
and stresses the importance of phenotypic variation in ecological stability. 

Our starting point is the Lotka--Volterra model~\cite{murray}, one of the most studied in population dynamics: $N$ species with abundances $\vec{A}$ compete as
\begin{align}
\vec{\dot{A}}=\vec{A}\bigl(\vec{\alpha}-\mat{D}\cdot\vec{A}\bigr), \label{eq:lv1}
\end{align}
where $\vec{\alpha}$ is a vector of birth rates and $\mat{D}$ a matrix of non-negative competition strengths~\footnote{Throughout, we imply elementwise multiplication of vectors and (rows or columns of) matrices by writing symbols next to each other. Dots denote tensor contractions.}; this is the \emph{competitive} (as opposed to predator-prey) flavor of the model.
If $\det{\mat{D}}\not=0$, Eq.~\eqref{eq:lv1} has a unique equilibrium $\vec{A_\ast}=\mat{D}^{-1}\cdot\vec{\alpha}$ of coexistence of all $N$ species. This equilibrium is feasible (i.e. $\vec{A_\ast}>\vec{0}$) if and only if $\vec{\alpha}$ lies in the positive span of the columns of $\mat{D}$. 

\begin{figure}[b]
\includegraphics{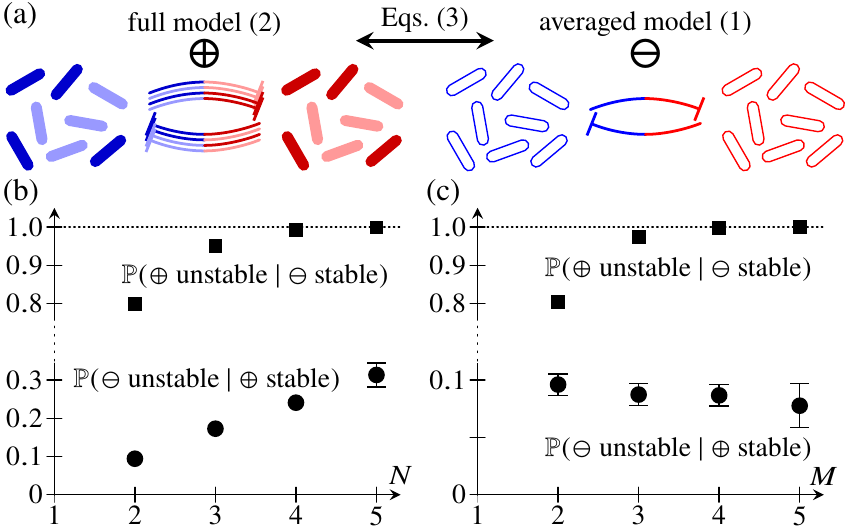}
\caption{Stability of communities with subpopulations. (a)~The consistency conditions~\eqref{eq:cc} relate the full model~\eqref{eq:lv2}, in which each species has two subpopulations (dark and light individuals) that interact with all subpopulations of every other species, to the corresponding averaged model~\eqref{eq:lv1} without subpopulations. (b)~Probability that a random stable equilibrium of the averaged model \eqref{eq:lv1}, $\ominus$, or the full model \eqref{eq:lv2}, $\oplus$, is unstable in the other model, as a function of the number of species, $N$. Probabilities were estimated from up to $10^7$ random systems. Error bars indicate $95\%$ confidence intervals larger than the plot markers. (c)~Corresponding results for models with $N=2$ species each having $M$ subpopulations, as a function of $M$.}
\label{fig1}
\end{figure}

If the $N$ species each have two subpopulations with respective abundances $\vec{B},\vec{C}$, then
\begin{align}
\vec{\dot{B}}=\vec{B}\bigl(\vec{\beta}\!-\!\mat{E}\!\cdot\!\vec{B}\!-\!\mat{F}\!\cdot\!\vec{C}\bigr),
\;\;\vec{\dot{C}}=\vec{C}\bigl(\vec{\gamma}\!-\!\mat{G}\!\cdot\!\vec{B}\!-\!\mat{H}\!\cdot\!\vec{C}\bigr),\label{eq:lv2} 
\end{align}
where $\vec{\beta},\vec{\gamma}$ are birth rates, and $\mat{E},\mat{F},\mat{G},\mat{H}$ are competition strengths~\cite{*[{For $N=2$, systems of this form have been considered in analyses of evolutionary stability, see e.g. }] [] cressman03}. The dynamics of the
sum $\vec{B}+\vec{C}$ derived from the subpopulation-resolving ``full'' system~\eqref{eq:lv2} are not of the averaged form~\eqref{eq:lv1}. However, to a coexistence equilibrium $(\vec{B_\ast},\vec{C_\ast})$ of Eq. \eqref{eq:lv2} we can 
associate an equilibrium $\vec{A_\ast}$ of Eq. \eqref{eq:lv1}, determined by the requirement that, at equilibrium, population sizes, births, and competition be equal \figref{fig1}{a}, i.e. that
\begin{subequations}
\label{eq:cc}
\begin{align}
\vec{A_\ast}&=\vec{B_\ast}+\vec{C_\ast},\qquad
\vec{\alpha A_\ast}=\vec{\beta B_\ast}+\vec{\gamma C_\ast},
\end{align}
and
\begin{align}
\vec{A_\ast}\mat{D}\vec{A_\ast}&=\vec{B_\ast}\left(\mat{E}\vec{B_\ast}+\mat{F}\vec{C_\ast}\right)+\vec{C_\ast}\left(\mat{G}\vec{B_\ast}+\mat{H}\vec{C_\ast}\right).
\end{align}
\end{subequations}
These consistency conditions are a property of the model, based on the interpretation of its terms. Given Eqs.~\eqref{eq:lv2}, they uniquely define an equilibrium $\vec{A_\ast}$ and an averaged model of the form~\eqref{eq:lv1}, and this $\vec{A_\ast}$ is an equilibrium of this averaged model, and feasible if $(\vec{B_\ast},\vec{C_\ast})$ is. 

We select random averaged and subpopulation-resolving systems by sampling model parameters from a uniform distribution~\footnote{See Supplemental Material at [url to be inserted], which includes \mbox{Refs.~\protect\cite{butler18,hinch,lidskii,linalg,grilli17,hardin60,*armstrong76,harms16,kussell05}}, for (i) a detailed description of the selection of random systems, (ii) details of the asymptotic calculations leading to Eqs.~\eqref{eq:ja} and \eqref{eq:ja0}, (iii)~details of the spectral analysis leading to Eq.~\eqref{eq:pp}, including a note on the stability of sums of stable matrices, (iv)~a discussion of models with explicit resource dynamics, based on those of Ref.~\cite{butler18}, and (v)~a brief discussion of the dynamics of antibiotic treatments.}, and analyze the stability of their coexistence equilibria by computing the eigenvalues of their Jacobians. \nocite{butler18,hinch,lidskii,linalg,grilli17,hardin60,armstrong76,harms16,kussell05} As the number of species $N$ increases, stable equilibria of the averaged model~\eqref{eq:lv1} are increasingly likely to be unstable in the full model~\eqref{eq:lv2} \figref{fig1}{b}. This is because the full model effectively has $2N$ species, and stable equilibria become increasingly rare as the number of species increases~\cite{may72,allesina12}. It is therefore all the more striking that, as $N$ increases, stable equilibria of the full model are also increasingly likely to be unstable in the averaged model \figref{fig1}{b}. The full model~\eqref{eq:lv2} can be extended to species with $M\geq 2$ subpopulations each, but increasing $M$ at fixed $N=2$ does not significantly affect the probability that a stable equilibrium of the full model destabilises in the averaged model \figref{fig1}{c}. 

These toy models thus show that implicit averaging of subpopulations leads to incorrect stability results, and hence underline their importance. Mathematically, this result is not fundamentally surprising: the determinant of the sum of two matrices is not the sum of their determinants, and so the linear relations between the Jacobians of the two systems resulting from Eqs.~\eqref{eq:cc} cannot be expected to lead to simple relations between their stability. 

We now specialize by taking phenotypic variation in microbial communities as an instance of subpopulation structure. It is useful to focus on one particular biological example: bacterial persisters~\cite{maisonneuve14,harms16}. Bacteria such as \emph{Escherichia coli} switch between a normal growth state and a persister state, in which they significantly suppress growth but are resilient to conditions of stress such as competition or exposure to antibiotics~\cite{harms16,andersson14,radzikowski17}. Infections can thus be difficult to treat even in the absence of genetic antibiotic resistance; for this reason, this phenotype has great biomedical relevance~\cite{harms16,andersson14}. 

Adding switching between normal cells and persisters to Eqs.~\eqref{eq:lv2} leads to a phenotype-resolving model,\begin{subequations}\label{eq:lv2s}
\begin{align}
\vec{\dot{B}}&=\vec{B}\bigl(\vec{\beta}-\mat{E}\cdot\vec{B}-\mat{F}\cdot\vec{C}\bigr)-\vec{\kappa}\vec{B}+\vec{\lambda C},\\
\vec{\dot{C}}&=\vec{C}\bigl(\vec{\gamma}-\mat{G}\cdot\vec{B}-\mat{H}\cdot\vec{C}\bigr)+\vec{\kappa}\vec{B}-\vec{\lambda C},
\end{align}
\end{subequations}
where $\vec{\kappa},\vec{\lambda}$ are rates of (stochastic) switching. This form has previously been used to study phenotypic switching of a single species without competition~\cite{balaban04,kussell05}. Since the rates are balanced, given an equilibrium of Eqs.~\eqref{eq:lv2s}, the consistency conditions \eqref{eq:cc} still define a corresponding ``averaged'' model \eqref{eq:lv1} without phenotypic variation.

\begin{figure}
\includegraphics{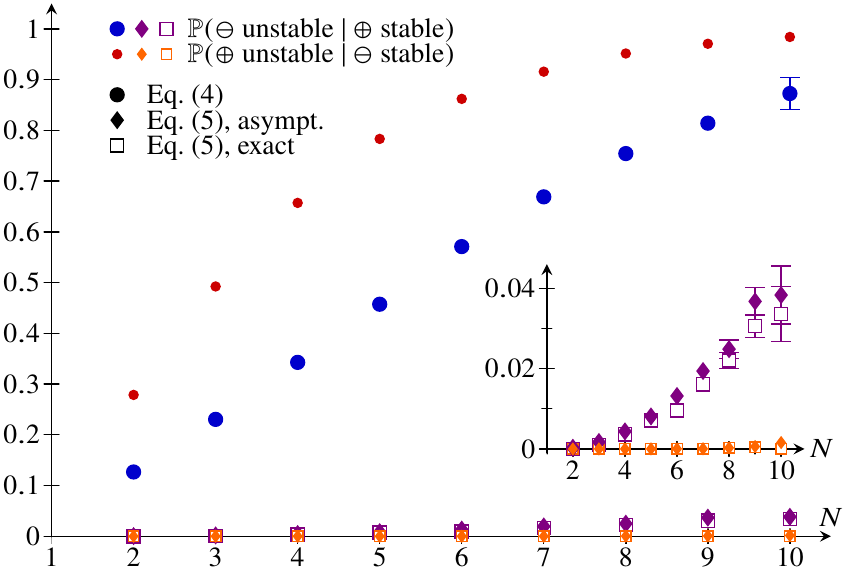}
\caption{Stability of microbial communities with phenotypic switching. Probability that a random stable equilibrium of a phenotype-resolving model $\oplus$ (large markers) or the corresponding averaged model~$\ominus$ (small markers) is unstable in the other model, as a function of the number of species, $N$. Different markers represent Eqs.~\eqref{eq:lv2s} and asymptotic and exact evaluations of Eqs.~\eqref{eq:lv2a2}. The inset focuses on low probabilities. Probabilities were estimated from up to $10^8$ random systems. Error bars indicate $95\%$ confidence intervals larger than the plot markers. Parameter value for Eqs.~\eqref{eq:lv2a2}: $\varepsilon=0.01$.}
\label{fig2}
\end{figure}

To analyze the effect of this switching, we again compare the stability of the full and averaged models. Steady states of Eqs.~\eqref{eq:lv2s} cannot in general be found in closed form. To sample random systems, we must therefore sample parameters indirectly~\cite{Note2}. With increasing number of species, random stable coexistence states of either this full model or the corresponding averaged model again become increasingly likely to be unstable in the other model \wholefigref{fig2}.

Eqs.~\eqref{eq:lv2s} do not however take into account the weak growth and competition of the persisters or the large separation of switching rates~\cite{balaban04}. Adding a small parameter $\varepsilon\ll 1$, we therefore modify Eqs.~\eqref{eq:lv2s} into\begin{subequations}\label{eq:lv2a2}
\begin{align}
\vec{\dot{B}}&=\vec{B}\bigl(\vec{\beta}-\mat{E}\cdot\vec{B}-\varepsilon\mat{F}\cdot\vec{C}\bigr)-\varepsilon\vec{\kappa}\vec{B}+\vec{\lambda C},\\
\vec{\dot{C}}&=\varepsilon\vec{C}\bigl(\vec{\gamma}-\mat{G}\cdot\vec{B}-\mat{H}\cdot\vec{C}\bigr)+\varepsilon\vec{\kappa}\vec{B}-\vec{\lambda C}.
\end{align}
\end{subequations}
Hence $\vec{B}$ are normal cells and $\vec{C}$ are persisters. For wild-type \emph{E. coli}, $\varepsilon\approx 10^{-5}$~\cite{balaban04}, but here, we take $\varepsilon=0.01$ for numerical convenience. We justify this below by confirming the numerical results by an asymptotic analysis of Eqs.~\eqref{eq:lv2a2}. A more intricate asymptotic separation of parameters arises in the \emph{hipQ} mutant of \mbox{\emph{E. coli}}~\cite{balaban04}, but we do not pursue this further.

The separation of growth and competition terms and switching rates in Eqs.~\eqref{eq:lv2a2} allows for steady states to be found by expansion in $\varepsilon$. Writing $\vec{B_\ast}=\vec{B_0}+\varepsilon\vec{B_1}+O\bigl(\varepsilon^2\bigr)$ and $\vec{C_\ast}=\vec{C_0}+\varepsilon\vec{C_1}+O\bigl(\varepsilon^2\bigr)$, we find $\vec{B_0}=\mat{E}^{-1}\cdot\vec{\beta}$ and $\vec{C_1}=\vec{\kappa B_0}/\vec{\lambda}$, but $\vec{B_1}=\vec{C_0}=\vec{0}$~\cite{Note2}. This is the expected asymptotic separation of the two population sizes: few cells are persisters, at least under laboratory conditions~\cite{balaban04}. This asymptotic solution enables direct sampling of all model parameters~\cite{Note2}.

To analyze the stability of equilibria of Eqs.~\eqref{eq:lv2a2}, we expand its Jacobian, $\mat{J_\ast}=\mat{J_0}+\varepsilon\mat{J_1}+O\bigl(\varepsilon^2\bigr)$, finding
\begin{align}
&\mat{J_0}=\left(\!\begin{array}{c|r}-\vec{B_0}\mat{E}&\vec{\lambda}\mat{I}\\\hline\mat{O}&-\vec{\lambda}\mat{I}\end{array}\!\right)\!,\; \mat{J_1}=\left(\!\begin{array}{r|c}-\vec{\kappa}\mat{I}&-\vec{B_0}\mat{F}\\\hline\vec{\kappa}\mat{I}&(\vec{\gamma}-\mat{G}\cdot\vec{B_0})\mat{I}\end{array}\!\right)\!,\label{eq:ja}
\end{align}
with $\mat{I}$ the identity and $\mat{O}$ the zero matrix~\cite{Note2}. The averaged model has Jacobian $\mat{K_\ast}=\mat{K_0}+\varepsilon\mat{K_1}+O\bigl(\varepsilon^2\bigr)$, with
\begin{align}
\mat{K_0}&=-\vec{B_0}\mat{E}, &\mat{K_1}&=\vec{B_0}\mat{E}\dfrac{\vec{\kappa}}{\vec{\lambda}}.\label{eq:ja0}
\end{align}
Since $\mat{J_0}$ is block-upper-triangular, its eigenvalues are those of $-\vec{\lambda}\mat{I}$, which are stable, and those of $-\vec{B_0}\mat{E}=\mat{K_0}$. Hence any unstable eigenvalues of $\mat{J_\ast}$ and $\mat{K_\ast}$ are equal to lowest order in the expansion. Equivalently, 
at $\varepsilon=0$, the full phenotype-resolving model~\eqref{eq:lv2a2} is stable if and only if the corresponding averaged model is stable. This result is not borne out however by numerics at finite $\varepsilon$: as $N$ increases, the probability that a random stable equilibrium of the full model is destabilized in the corresponding averaged model still increases \wholefigref{fig2}, although the probability is reduced compared to the previous case. Much more strikingly, the probability that a stable equilibrium of the averaged model is destabilized in the full model is vastly reduced \wholefigrefi{fig2}. This is all the more surprising as we argued earlier that the opposite behavior was to be expected since larger systems are more likely to be unstable. We have also sampled exact equilibria of Eqs.~\eqref{eq:lv2a2}, similarly to our analysis of Eqs.~\eqref{eq:lv2s} above, yielding results in qualitative agreement with those based on the asymptotic equilibria~\wholefigrefi{fig2}. This justifies basing the detailed analysis of the destabilization mechanism on the asymptotic results.

\begin{figure}
\includegraphics{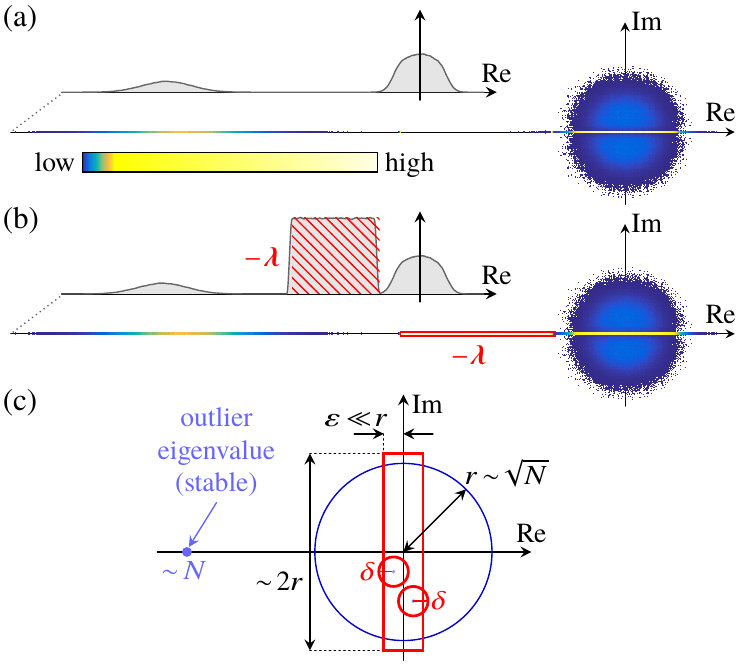} 
\caption{Eigenvalue distributions of the full and averaged models for $\varepsilon\ll 1$. (a)~Eigenvalue distribution, for $N=10$, of the Jacobian $\mat{K_\ast}$ of Eq.~\eqref{eq:lv1}. Histogram obtained from $10^5$ random systems. Inset shows distribution of real eigenvalues. (b)~Corresponding plot for the Jacobian $\mat{J_\ast}$ of Eqs.~\eqref{eq:lv2a2}. Parameter value: $\varepsilon=0.01$. Boxes on the real axis and in the inset indicate the real eigenvalues of $-\vec{\lambda}\mat{I}$. (c)~Eigenvalue distribution of $\mat{K_0}$ with a stable outlier eigenvalue and a circular core of radius $\smash{r\sim\sqrt{N}}$. The stability of eigenvalues $\nu_0$ with $|\mathrm{Re}(\nu_0)|\lesssim\varepsilon\ll r$ can be affected by higher-order terms. The average distance between eigenvalues is $\delta=O(1)$.}
\label{fig3}
\end{figure}

To explain these surprising results, we analyze the spectra of the Jacobians in more detail.  With the exception of a single outlier eigenvalue that is large and negative, the eigenvalues of $\mat{K_\ast}$ lie approximately within a circle~\figref{fig3}{a}, as expected from the circular law of random matrices~\cite{tao10}. The spectral distribution of $\mat{J_\ast}$ is the sum of this distribution and the (uniform) distribution of eigenvalues of $-\vec{\lambda}\mat{I}$ \figref{fig3}{b}. The outlier eigenvalue can be analyzed in great generality~\cite{tao13}, but heuristics suffice here: denoting by $-\mu_0<0$ the mean of the distribution of entries of $\mat{K_0}$ and neglecting correlations between entries, each row of $\mat{K_0}$ has approximate sum $-N\mu_0$, and so $\mat{K_0}$ has an approximate eigenvector $\mathbf{1}=(1,1,\dots,1)$ with eigenvalue $-N\mu_0<0$, as argued in the Supplemental Material of \citeref{allesina12}. The other eigenvalues of $\mat{K_0}$ \figref{fig3}{c} are uniformly distributed on a disk of radius $r\sim\sqrt{N}$ for $N\gg 1$ by the circular law~\cite{tao10}. (Hence, by the Perron--Frobenius theorem~\cite{* [] [{ Chap. 1.3, pp. 44--56, Chap. 2.5, pp. 100--112, and Chap. 8.2, pp. 495--503.}] linalg}, the outlier eigenvalue is indeed real.) An eigenvalue $\nu_0$ of $\mat{K_0}$ has $|\mathrm{Re}(\nu_0)|\lesssim\varepsilon$ with probability $\varpi\sim (\varepsilon r)\big/r^2\sim\varepsilon\big/\sqrt{N}$, i.e. $\varpi = c\varepsilon\big/\sqrt{N}$, for some $c=O(1)$. The average distance $\delta$ between eigenvalues is determined by $N\delta^2\sim r^2$, so $\delta=O(1)$ and the eigenvalues of $\mat{K_0}$ are pairwise different at order $O(1)$~\figref{fig3}{c}. Hence, if $\nu_0$ is an eigenvalue of $\mat{K_0}$, then $\mat{K_\ast}$ has an eigenvalue $\nu_\ast=\nu_0+O(\varepsilon)$~\cite{Note2}. Thus $\nu_\ast$ is stable if either (i)~$\mathrm{Re}(\nu_0)<0$ and $|\mathrm{Re}(\nu_0)|\gtrsim\varepsilon$ or (ii) $|\mathrm{Re}(\nu_0)|\lesssim\varepsilon$ and the small real part of $\nu_\ast$ is stabilized by $\mat{K_1}$. By definition, $|\mathrm{Re}(\nu_0)|\lesssim\varepsilon$ with probability $\varpi$, so (i) occurs with probability $(1-\varpi)/2$. Let $q$ denote the probability of stabilization by $\mat{K_1}$ in case (ii). Summing over the $N-1$ non-outlier eigenvalues of~$\mat{K_0}$, the probability $p=\mathbb{P}(\mat{K_\ast}\,\text{stable})$ is
\begin{align}
p&=\sum_{k=0}^{N-1}{\binom{N-1}{k}\varpi^kq^k\left(\dfrac{1-\varpi}{2}\right)^{N-1-k}}\nonumber\\
&=\left(\dfrac{1}{2}+\dfrac{c(2q-1)\varepsilon}{2\sqrt{N}}\right)^{N-1}\sim\dfrac{\exp{\left[c(2q-1)\varepsilon\sqrt{N}\right]}}{2^{N-1}},
\end{align}
for $N\gg 1$, using the binomial theorem and $(1+1/x)^x\sim\mathrm{e}^x$ for $x\gg 1$. A similar expression determines $p'=\mathbb{P}(\mat{J_\ast}\,\text{stable})$, with $q$ replaced $q'$. Eq.~\eqref{eq:ja} shows that $\mat{J_1}$ acts on $\mat{K_0}$ as the negative definite matrix $-\vec{\kappa}\mat{I}$, so $q'>q$~\cite{Note2}. It follows that \begin{align}
\dfrac{p'}{p}\sim\exp{\left[2c(q'-q)\varepsilon\sqrt{N}\right]}\rightarrow\infty\quad\text{as }N\rightarrow\infty,\label{eq:pp}
\end{align}
confirming the trend in \textwholefigref{fig2}: the full model is much more likely to be stable than the averaged model.

Hence, switching to a rare phenotype such as persisters can enhance the stability of a community. The detailed analysis of Eqs.~\eqref{eq:lv2a2} above has emphasized that this effect relies on both the spectral distribution (which allows terms beyond leading order to change the stability of the system) and the detailed structure of the system (which can suppress or enhance this mechanism). By contrast, switching to an abundant phenotype destabilizes the community: the introduction of such a phenotype essentially increases the effective number of species, which is destabilizing~\cite{may72,allesina12}. Reference~\cite{butler18} recently introduced a family of models with explicit resource dynamics for which any feasible equilibrium is stable. Switching to an abundant phenotype also destabilizes a phenotype-resolving version of the model of \citeref{butler18} provided that the difference in switching rates is large enough~\cite{Note2}, confirming that this effect is generic. These conclusions are therefore likely to be relevant for the stability of microbial communities such as the microbiome, for which competitive interactions are known to play an important, stabilizing role~\cite{coyte15}.

Linear stability analysis cannot elucidate the effect of large perturbations on coexistence. These might arise from antibiotic treatments, which bacterial communities can survive by forming persisters~\cite{harms16}. In the final part of this Letter, we therefore explore such perturbations numerically. 
Rather than modelling the dynamics of antibiotic treatment in detail~\cite{Note2}, we suppose that it reduces the abundances of both normal cells and persisters, and we ask: does the community converge back to coexistence, or do some species disappear from the community? Do the answers from the averaged and full models differ? To answer these questions, we reconsider the exact stable equilibria of the full model~\eqref{eq:lv2a2} that are also stable in the averaged model, and evolve both systems from consistent random small initial conditions using the stiff solver \texttt{ode15s} of \textsc{Matlab} (The MathWorks, Inc.).

\begin{figure}
\includegraphics{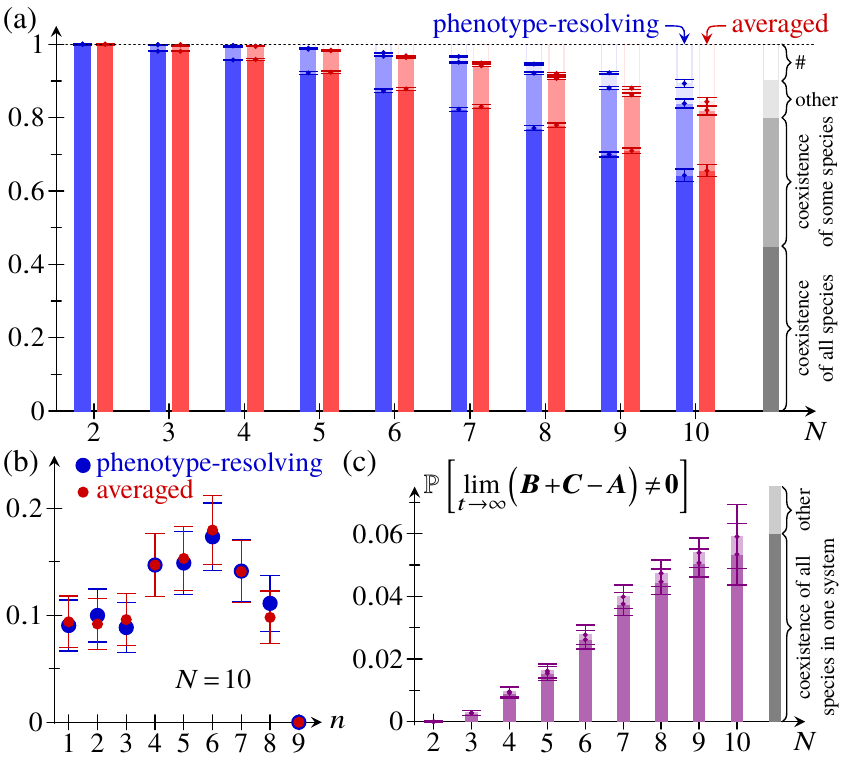}
\caption{Stability of microbial communities to large perturbations. (a)~Distribution of possible long-time behaviors after large perturbations of the full model~\eqref{eq:lv2a2} and the corresponding averaged model as a function of $N$. Possible behaviors are (1)~convergence back to coexistence of all species, (2) convergence to coexistence of $n<N$ species, or (3) non-convergence to a steady state. In some cases (\#), numerical solution failed. Probabilities were estimated from up to $10^4$ random systems. (b)~Distribution of the number $n$ of remaining species for outcome (2), for $N=10$ and for both the full and averaged models. (c)~Probability that the full and averaged models converge to different coexistence states, given that both models converge to some coexistence state. The contribution from systems for which one model converges to coexistence of all species is highlighted. Error bars indicate $95\%$ confidence intervals.}
\label{fig4} 
\end{figure}

\textfigref{fig4}{a} shows the distribution of possible outcomes: (1)~convergence back to coexistence of all $N$ species, (2) convergence to a new coexistence state of $n<N$ species, and (3)~convergence to a limit cycle. (The trivial equilibria $\vec{A}=\vec{0}$ and $\vec{B}=\vec{C}=\vec{0}$ of the averaged and full models are clearly unstable, so $1\leq n<N$.) The probability of outcome (2) increases with $N$~\figref{fig4}{a} and the distribution of $n$ in~\textfigref{fig4}{b} shows that $n\gtrapprox N/2$ is somewhat more likely than $n\lessapprox N/2$. Thus if the whole community does not survive, then 
at least half does. The averaged and full persister models give comparable outcome distributions. This does not contradict our earlier result that a rare phenotype stabilizes the community since here, we only consider states stable in both the averaged and full models so that results can be compared meaningfully. In fact, as $N$ increases, individual realizations of full and averaged models are increasingly likely to give different outcomes~\figref{fig4}{c}. In most cases, outcomes differ because the system converges back to coexistence of all species in one model only; in a small number of cases, different species die out in the full and averaged models~\figref{fig4}{c}. These observations thus extend our results for linear perturbations to large perturbations. 

Here, we have revealed the strong effects of subpopulation structure on the stability of competing microbial communities and the surprising stabilizing effect of stochastic switching to a rare phenotype. Very recently, Ref.~\cite{maynard19} similarly emphasized the stabilizing effect of phenotypic variation using a different model, without phenotypic switching. While the competitive interactions considered here are important in systems such as the human microbiome~\cite{coyte15}, future work will need to explore the phenotypic structure in more detail. The interaction structure of ecological communities without phenotypic variation is known to affect their stability~\cite{allesina12,grilli16}, but these and related studies, in the spirit of May's seminal work~\cite{may72}, are based on the analysis of random Jacobians. By contrast, here, we could not avoid specifying explicit dynamical systems, since we had to establish a correspondence between full and averaged models (and indeed our analysis has shown that the details of the model structure can matter here). In particular, this prepends the question of feasibility to the question of stability. This question of feasibility can be treated in a statistical sense~\cite{grilli17b,gibbs18} but cannot be eschewed in general. Nonetheless, our analysis only relying on generic properties of the spectral distribution suggests that our conclusions apply not only to competitive, but also to more general interactions.

\begin{acknowledgments}
The authors gratefully acknowledge support from Engineering and Physical Sciences Research Council Established Career Fellowship EP/M017982/1 and Gordon
and Betty Moore Foundation Grant 7523 (both to R.E.G.), a Herchel Smith Postdoctoral Research Fellowship (N.M.O.), and Magdalene College, Cambridge (Nevile Research Fellowship to P.A.H.).
\end{acknowledgments}

\bibliography{phe}
\end{document}